\documentclass[reprint,amsmath,amssymb,aps, twocolumn]{revtex4-2}

\usepackage{hyperref}
\usepackage{comment}
\hypersetup{colorlinks,urlcolor=blue}
\usepackage{graphicx}
\usepackage[percent]{overpic}
\usepackage{dcolumn}
\usepackage{bm}
\usepackage{ulem}
\usepackage{physics}
\usepackage{lipsum}
\usepackage{appendix}
\usepackage[T1]{fontenc}

\newcommand{\edit}[1]{\textcolor{black}{#1}}

\newcommand{\ham}{\mathcal{H}}

\begin{document}
\preprint{APS/123-QED}
\title{Dynamics of Spin Helices in the one-dimensional $XX$ Model}
\author{Darren Pereira}
\author{Erich J. Mueller}
\affiliation{Laboratory of Atomic and Solid State Physics, Cornell University, Ithaca, New York 14853, USA}

\begin{abstract}
Motivated by cold-atom experiments and a desire to understand far-from-equilibrium quantum transport,  we analytically study the dynamics of spin helices in the one-dimensional $XX$ model. 
We use a Jordan-Wigner transformation to map the spin chain onto a non-interacting Fermi gas with simple equations of motion. The resulting dynamics are nontrivial, however, as the spin-helix initial condition corresponds to a highly nonequilibrium distribution of the fermions. 
We find a separation of timescales between the in-plane and out-of-plane spin dynamics.  We gain insights from analyzing the case of a uniform spin chain and from a semiclassical model.  One of our key findings is that the spin correlation functions decay as $t^{-1/2}$ at long time, in contrast to the experimentally observed exponential decay.
\end{abstract}

\date{\today}
\maketitle

\section{Introduction}
Much of our modern understanding of highly entangled quantum matter
has come from the study of spin chains. Exemplary phenomena include topological order \cite{AKLTPRL1987, KitaevUFN2001} and many-body localization \cite{BaskoAnnPhys2006}.  Spin chain models were also pivotal in developing numerical techniques such as the density matrix renormalization group  \cite{WhitePRL1992}. Cold-atom experiments can now controllably implement spin chain models, allowing experimental investigation of their properties \cite{JepsenNature2020, JepsenPRX2021}. While the models are old, the cold-atom realizations naturally lend themselves to experimentally studying novel scenarios. For example, Jepsen \textit{et al.} \cite{JepsenNature2020} initialized a spin chain in a helix and then observed how it evolved under the one-dimensional  $XX$ or $XXZ$ model Hamiltonians. Studies of  such highly non-equilibrium dynamics are relatively rare. 
Here, we model this spin helix dynamics experiment in the case of the $XX$ model.

The $XX$ model is an extreme limit of the anisotropic Heisenberg model, where the two in-plane components of the exchange coupling are equal and the out-of-plane coupling vanishes: 
\begin{equation}
    \ham = \sum_j J \left[\sigma_j^x \sigma_{j+1}^x + \sigma_j^y \sigma_{j+1}^y\right]. \label{eq:XXModelSpins}
\end{equation}
Here $j$ labels the site on the chain, $\sigma_j^\mu$ is the Pauli matrix for the $\mu$-th component ($\mu$ = $x$, $y$), and $J$ is the strength of the coupling. The importance of this model comes from 
the fact that it can be mapped onto a gas of non-interacting spinless fermions. This mapping nominally reduces all static and dynamic calculations to exercises in single-particle quantum mechanics.  Related transformations have become ubiquitous in condensed matter physics \cite{FradkinPRL1989, WangPRB1991, BatistaPRL2001, RossiniPRB2004, KikoinPRB2005, DerzhkoChapter, WenQFT2004, FradkinQFT2013}. We use this mapping to model the experiment of Jepsen \textit{et al.} \cite{JepsenNature2020}. 

In that experiment, a gas of (bosonic) $^{7}$Li atoms is trapped in an ensemble of 1D optical lattices, nominally with one particle per site. Strong interactions suppress  hopping, and 
the $\sigma_j^z=\pm 1$ spin states correspond to different
hyperfine states of the localized atoms.
Superexchange leads to nearest-neighbor spin-spin interactions, whose strength can be tuned with a Feshbach resonance \cite{TiesingaRMP2010}.  This allows the implementation of  Eq.~\eqref{eq:XXModelSpins} or other spin models. The spin chain is initialized in a helix with $\langle \sigma_j^z\rangle=-\cos(Qj+\phi)$ and  $\langle \sigma_j^x\rangle=\sin(Qj+\phi)$, where $Q=2\pi/\lambda$ is the wave vector of the helix and $\phi$ is its phase.

In the language of fermions, the spin-helix initial state is highly unusual.  We diagonalize the single-particle density matrix and show that the state is made up of several bands with fractional occupation.  Starting from this initial state, we calculate the dynamics of spin correlations. Their time dependence is expressed in terms of a sum of Bessel functions, from which the long-time asymptotics are readily extracted. We compare these exact results to semiclassical and long-wavelength approximations. This leads to a physical picture in terms of spin precession and quantum spin diffusion, with the correlations decaying as $t^{-1/2}$.

The experiment instead found that this decay was exponential and used the $Q$ dependence of the time constant to distinguish between models with different transport regimes   (ballistic, superdiffusive, diffusive, and subdiffusive).  
\edit{The discrepency is likely due to the presence of empty sites in their lattice \cite{JepsenNature2020} or inhomogeneities in their optical potentials or magnetic fields. Although we do not model it here, we suspect the discrepancy originates from the empty sites in particular. Initial simulations modeling the effect of such holes have been performed in Ref.~\cite{JepsenNature2020}, and a disagreement with the experiment is already present there.}
The accurate quantum simulation of iconic spin models in the future will require addressing such \edit{experimental} complications.

The outline of the paper is as follows. In Sec.~\ref{model}, we  describe the Jordan-Wigner transformation and examine the spin helix from the fermion perspective.  We analyze the single-particle density matrix, describing the normal modes.  In Sec.~\ref{dynamics}, we detail the formalism for calculating spin correlations, and in Sec.~\ref{sec:Results}, we present our results.  In addition to calculating the exact dynamics, we present a long-wavelength approximation and compare to the semiclassical dynamics. 
Finally, we compare our results to the motivating experiment of Jepsen \textit{et al.} \cite{JepsenNature2020} in Sec.~\ref{sec:ExpComp} before giving a brief summary and outlook in Sec.~\ref{sec:conclusion}.

\section{Setup}\label{model}

\subsection{Jordan-Wigner Transformation}

The $XX$ model in Eq.~(\ref{eq:XXModelSpins}) is diagonalized by mapping the Pauli spin operators onto fermionic operators via the Jordan-Wigner transformation \cite{JordanWigner, LiebAnnPhys1961, ColemanMBP2004},
\begin{align}\label{sigp}
    \sigma_j^+ &= 2e^{-i \pi \sum_{k=1}^{j-1} n_k} a_j^\dagger, \\
	\sigma_j^- &= 2e^{i \pi \sum_{k=1}^{j-1} n_k} a_j, \\\label{sigz}
	\sigma_j^z &= 2n_j -1,
\end{align}
where $a_j$ and $a_j^\dagger$ are fermion annihilation and creation operators, and the fermion occupation number operator is $n_j = a_j^\dagger a_j$. Note the factors of two in Eqs.~(\ref{sigp})-(\ref{sigz}) due to writing the transformation in terms of Pauli operators. Under this transformation, Eq.~\eqref{eq:XXModelSpins} 
becomes
\begin{eqnarray}
    \ham &=& 2J \sum_j \left[a_j^\dagger a_{j+1} + a_{j+1}^\dagger a_j\right] \label{eq:XXModelFermions}\\
    &=& \sum_k \epsilon(k) b_k^\dagger b_k,
\end{eqnarray}
where $\epsilon(k)= 4J\cos(k)$ is the fermion dispersion, 
$b_k =L^{-1/2} \sum_j e^{-ikj} a_j$, and $L$ is the number of sites. For our analytic calculations we take $L\to\infty$.  For our numerics we use a  large  value of $L$ and show results far from the boundary.

\subsection{Initial Conditions}\label{ic}

In  Ref.~\cite{JepsenNature2020}, the spins are chosen to form a helix in the $x$-$z$ plane,
\begin{eqnarray}
    \ket{\psi} &=& \prod_j \left[A_{j,\uparrow} \ket{\uparrow}_j + A_{j,\downarrow} \ket{\downarrow}_j \right], \label{eq:SpinState}
\end{eqnarray}
where
\begin{equation}
A_{j,\uparrow}=\sin(\frac{Qj+\phi}{2}),\qquad
A_{j,\downarrow}=
\cos(\frac{Qj+\phi}{2}),
\end{equation}
and where, as previously introduced, $Q = \frac{2\pi}{\lambda}$ is the helix wavevector and $\phi$ is the initial phase of the helix. This models a helix which satisfies $\langle \sigma_j^z\rangle=-\cos(Qj+\phi)$ and $\langle \sigma_j^x\rangle=\sin(Qj+\phi)$. In the fermion picture,
we calculate the single-particle density matrix,
\begin{eqnarray}
    \rho_{lm} &=& \expval{a_l^\dagger a_m}= \expval{\sigma_l^+ \sigma_m^- \prod_{i=l+1}^{m-1} \sigma_i^z }\\
    &=& \expval{\sigma_l^+}\expval{\sigma_m^-} \prod_{i=l+1}^{m-1} \expval{\sigma_i^z },\label{rhospin}
\end{eqnarray}
where the product is over all $i$ between $l$ and $m$, exclusively, and we used that the spins on different sites are uncorrelated. 
The diagonal elements are 
$\rho_{ll}= |A_{l,\uparrow}|^2$, and for $l<m$ the off-diagonal ones are
\begin{equation}
\rho_{lm}=	A_{m, \uparrow} A_{m, \downarrow}^* A_{l, \downarrow} A_{l, \uparrow}^*  \left(\prod_{i=l+1}^{m-1} \left(\abs{A_{i, \downarrow}}^2 - \abs{A_{i, \uparrow}}^2\right)\right).
\end{equation}
For $m<l$, one uses $\rho_{lm}=\rho_{ml}$.

\begin{figure}[t!]
    \centering
    \includegraphics[width=0.9\linewidth]{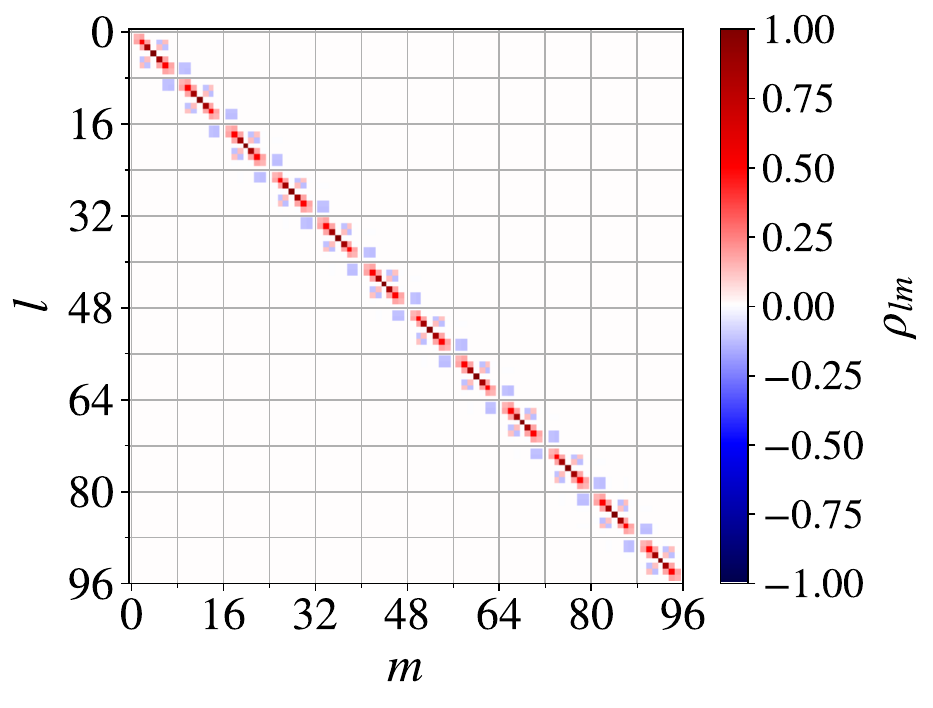}
    \caption{Density matrix $\rho_{lm} = \expval{a_l^\dagger a_m}$ [Eq.~\eqref{rhospin}] for the fermionic representation of a spin helix with wavelength $\lambda = 8$, phase $\phi=0$, and chain length $L = 96$. $l$ and $m$ are indices for the sites of the 1D chain. \edit{Open boundary conditions are used for this visualization.}}
    \label{fig:DensityMatrix}
\end{figure}

An example of the matrix elements of the density matrix $\rho_{lm}$ is shown in Fig.~\ref{fig:DensityMatrix} for a chain of length $L=96$, a spin helix of wavelength $\lambda=8$, phase $\phi=0$, \edit{and open boundary conditions}.   When $\lambda$ is an integer, the density matrix \edit{in the thermodynamic limit} is periodic, satisfying $\rho_{lm}=\rho_{l+\lambda,m+\lambda}$.  It is also largely local, with the  off-diagonal elements falling off rapidly.  This fall-off can be understood as a consequence of the  product  of $\langle \sigma_i^z\rangle$ terms in Eq.~(\ref{rhospin}). Since there are sites where $|\langle\sigma_i^z\rangle|\neq 1$, this product  always has a magnitude less than 1, leading to exponentially falling correlations.   Moreover, if there is a site where $\langle \sigma_j^z\rangle=0$, the density matrix has no nonzero elements with $|l-m|>\lambda$.

The structure when $\lambda$ is not an integer is more complicated. A rational $\lambda=p/q$  (where $p$ and $q$  are relatively prime) yields a density matrix with periodicity $p$.   Irrational $\lambda$ gives a quasiperiodic behavior.  Regardless, 
the elements $\rho_{lm}$ are very small when $|l-m|\gg\lambda$.  We will restrict our analysis to the case of integer $\lambda$ and leave exploration of  the   more general cases to the future.

The initial state also  contains pair correlations, which are encoded in $\Delta_{lm}=\langle a_l a_m\rangle$.  For $l<m$,
\begin{equation}
\Delta_{lm}=A_{m,\uparrow}^* A_{m, \downarrow} A_{l, \uparrow}^* A_{l, \downarrow} \left(\prod_{i=l+1}^{m-1} \abs{A_{i,\downarrow}}^2 -\abs{A_{i,\uparrow}}^2 \right), \label{eq:PairCorMat}
\end{equation}
and $\Delta_{ml}=\Delta_{lm}^*$.  
The diagonal elements vanish, $\Delta_{ll}=0$.
For states where the spins lie in the $x$-$z$ plane, $\Delta$ is real, and the off-diagonal elements of $\Delta$ are equal to the off-diagonal elements of $\rho$.
 
\subsection{Bloch Bands for an Integer-Wavelength Spin Helix}

To better understand the initial conditions, we diagonalize the density matrix, extracting the normal modes and their occupation numbers:
\begin{equation}
\rho_{lm}=\sum_\alpha f_\alpha (v^{(\alpha)}_l)^* v^{(\alpha)}_m.
\end{equation}
Here, $f_{\alpha}$ is the number of fermions in mode $\alpha$, with wavefunction 
$v^{(\alpha)}_m$.  These solve the eigenvalue problem
\begin{equation}
    \sum_m \rho_{lm} v_m^{(\alpha)} = f_{\alpha} v_l^{(\alpha)}. \label{eq:RhoEigenvalue}
\end{equation}
In a thermal ensemble of non-interacting fermions, the $v_l^{(\alpha)}$ are plane waves and  the occupation numbers correspond to a Fermi distribution.  As we show  below, the helix initial condition corresponds to a very different structure.

As previously stated, we specialize to the case where 
the wavelength $\lambda$ of the helix is an integer, and \edit{in the thermodynamic limit} $\rho_{lm} = \rho_{l+\lambda,m+\lambda}$. 
We then use 
 Bloch's theorem
 to write the eigenvectors in the form
$v_m^{(\alpha)} = e^{iqm} \phi_m^{(\alpha)}$, where $\phi_m^{(\alpha)}$ satisfies $\phi_{m + \lambda}^{(\alpha)} = \phi_m^{(\alpha)}$. Periodicity in $\lambda$ implies that $q \in \left[-\frac{\pi}{\lambda}, \frac{\pi}{\lambda} \right)$. Writing $m = n + s\lambda$, the eigenvalue equation \eqref{eq:RhoEigenvalue} becomes
\begin{equation}
    \sum_n \Gamma_{ln}(q) \phi_n^{(\alpha)} = f_{\alpha}(q) \phi_l^{(\alpha)}
\end{equation}
for the $\lambda \times \lambda$ matrix
\begin{eqnarray}
    \Gamma_{ln}(q) &=& \sum_s \rho_{l,n+s\lambda} e^{iq(n-l+s\lambda)}\\
    &=& \left(\rho_{ln}+\frac{\rho_{l,n+\lambda}e^{iq\lambda}}{1-\chi e^{iq\lambda}}+\frac{\rho_{l,n-\lambda}e^{-iq\lambda}}{1-\chi e^{-iq\lambda}}\right)e^{iq (n-l)}\nonumber,
\end{eqnarray}
where $0\leq l,n<\lambda$, and 
$\chi=\prod_{j=1}^\lambda (|A_{j,\downarrow}|^2-|A_{j,\uparrow}|^2)$.
The matrix
$\Gamma_{ln}(q)$ is numerically diagonalized to obtain occupation numbers for $\lambda$ separate ``bands'' as a function of wavevector $q$. 

\begin{figure}[t!]
    \centering
    \includegraphics[width=0.9\linewidth]{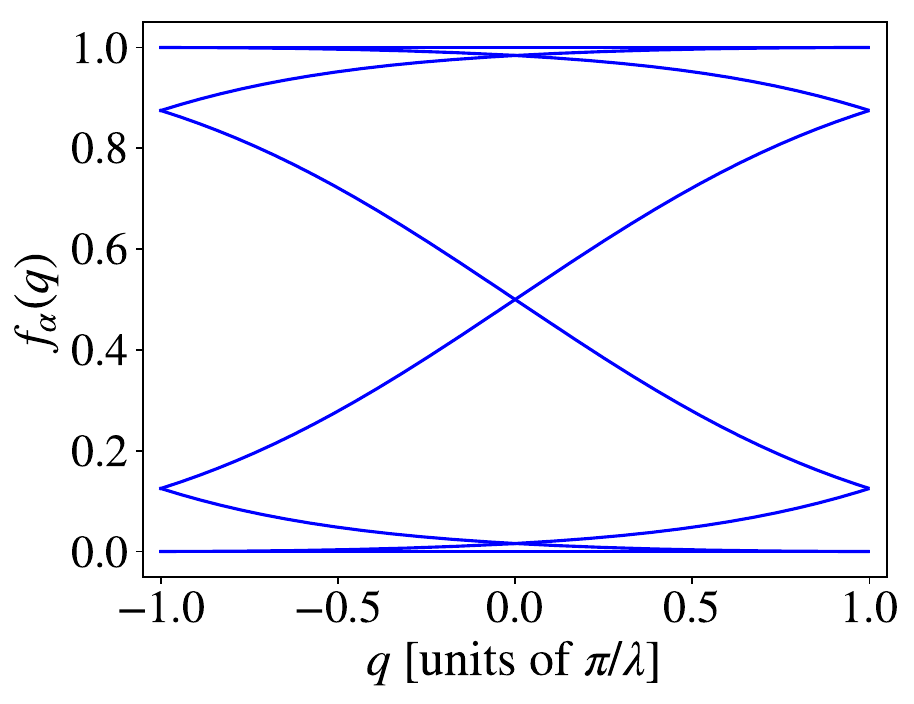}
    \caption{Occupation number bands for the 
Jordan-Wigner fermions
corresponding to the spin-helix initial condition.  These  are the eigenvalues of the
    density matrix $\rho_{lm}$ [Eq.~\eqref{rhospin}].  Here the helix has wavelength  $\lambda=8$ and phase $\phi=0$.}
    \label{fig:DensityBands}
\end{figure}

Figure \ref{fig:DensityBands}
shows the bands for $\lambda=8$ and $\phi=0$. In this particular case there is one band for which the occupation is uniformly 0 and another for which it  is uniformly 1.  These are a consequence of the choice of helix phase, $\phi=0$. For this $\phi$, there are sites where the spins are pointing in the $+z$ and $-z$ directions.  The fermions on those sites are localized and give rise to the uniformly filled or empty bands.  For arbitrary $\phi$, one instead finds a single state at $q=0$ for which $f_\alpha=1$, and a single state  for which $f_\alpha=0$.  This latter state is found at $q=\pi/\lambda$ if $\lambda$ is odd or $q=0$ if $\lambda$ is even. The presence of a fully occupied state and a fully empty state is a generic feature of any classical spin configuration, not just helices:  in complete generality, $v_m^{(1)}=A_{m,\uparrow}/A_{m,\downarrow}$ and $v_m^{(0)}=(-1)^m A_{m,\downarrow}/A_{m,\uparrow}$ are eigenvectors of Eq.~(\ref{eq:RhoEigenvalue}) with eigenvalues 1 and 0, respectively.  

One interesting feature of the occupation spectrum shown in Fig.~\ref{fig:DensityBands} is that there are no ``gaps." Given any value of $f_\alpha$, there is always a state with that occupation.  This is very different from a thermal  distribution in a multi-band model, where the energy gaps give rise to forbidden values of $f_\alpha$.
One can also observe a number of symmetries in Fig.~\ref{fig:DensityBands}.  When $\lambda$ is even, the helix is invariant under translation by half a wavelength  combined with spin reversal.  In the occupation spectrum this symmetry leads to a mapping $f_\alpha \to 1-f_\alpha$.    Reflection of the spins  through the $x$-$z$ plane maps  $q \to -q$.  The helix is invariant under this transformation, and hence Fig.~\ref{fig:DensityBands} has reflection symmetry.

\section{Formalism}\label{dynamics}
\subsection{Dynamics within the Fermionic Picture} \label{subsec:FermionDynamics}

Within the Heisenberg picture, the dynamics of the fermion field operators is trivial,
\begin{eqnarray}
b_k(t)=e^{-i\epsilon(k) t} b_k, \label{eq:FermionWaveTime}\\
a_j(t)=\sum_m i^{j-m}J_{m-j}(4Jt) a_m, \label{eq:FermionAnnTime}
\end{eqnarray}
where $b_k=b_k(0)$ and $a_j=a_j(0)$ are the zero-time operators, $J_\nu(x)$ is the Bessel function of the first kind of order $\nu$, and Eq.~\eqref{eq:FermionAnnTime} is the Fourier transform of Eq.~\eqref{eq:FermionWaveTime}.  Thus we can calculate any fermion correlation functions at time $t$ in terms of the correlations at time $0$, which were described in Sec.~\ref{ic}.  For example, the occupation of site $j$ is 
\begin{equation}
    \expval{n_j(t)} = \sum_{lm} \rho_{lm} i^{l-m} J_{l-j}(4Jt) J_{m-j}(4Jt). \label{eq:OccNum}
\end{equation}
Due to the structure of the density matrix, this sum is dominated by terms where $l$ and $m$ are close together.
Consequently, the dynamics at a given site $j$ are mainly influenced by the interference of fermions that propagate from closely-spaced sites on the chain.

\subsection{Spin Observables} \label{subsec:SpinObservables}

Some spin correlations can be simply expressed in terms of the fermion degrees of freedom and are readily calculated.  For example,
\begin{equation}
    \expval{\sigma_j^z(t)} = 2\expval{n_j(t)} -1 \label{eq:sigmazExpression}
\end{equation}
can directly be calculated from Eq.~(\ref{eq:OccNum}).  Others, such as $\langle \sigma_x(t)\rangle$, are highly non-local in the fermions.  A correlation function which is both easy to calculate and informative about the   transverse spin degrees of freedom is 
\begin{equation}
    E^{-+}_j(t) = \expval{\sigma_j^-(t) \sigma_{j+1}^+(t)}=4\langle a_{j+1}^\dagger(t) a_j(t) \rangle.
\end{equation}
The real and imaginary parts of $E_j^{-+}$ tell us about the relative alignment of the in-plane components of the spins on neighboring sites.
 $\Re\{ E^{-+}_j\} = \langle \sigma_j^x\sigma_{j+1}^x\rangle+\langle\sigma_j^y \sigma_{j+1}^y\rangle$ is the dot-product of the in-plane component of neighboring spins and is proportional to the local energy density.  When the spins are nearly aligned, it effectively measures the magnitude of the in-plane spin components. The other quadrature, $\Im \{E^{-+}_j\} = \langle  \left(
\vec \sigma_j\times \vec \sigma_{j+1}
 \right) \cdot \hat z\rangle$, measures the twisting (misalignment) of the in-plane spin components.  In terms of the fermion density matrix,
\begin{equation}
    E^{-+}_j(t) = 4\sum_{lm} \left[\rho_{lm} - \delta_{lm}\right] i^{l-m-1} J_{l-j-1}(4Jt) J_{m-j}(4Jt). \label{eq:E-+Expression}
\end{equation}
Similar expressions hold for the correlator
\begin{align}
    E^{--}_j(t) &= \expval{\sigma_j^-(t) \sigma_{j+1}^-(t)}=4\langle a_{j+1}(t) a_j(t) \rangle\\
&= 4\sum_{lm} \Delta_{lm} i^{2j-l-m+1} J_{l-j}(4Jt) J_{m-j-1} (4Jt). \label{eq:E--Expression}
\end{align}
The real and imaginary parts of $E_j^{--}$ are related to the in-plane quadrupolar alignment of neighboring spins:
$\Re\{ E^{--}_j\}=
\langle \sigma_j^x \sigma_{j+1}^x\rangle-
\langle \sigma_j^y\sigma_{j+1}^y\rangle$ and $\Im\{ E^{--}_j\}=-  (\langle \sigma_j^x \sigma_{j+1}^y\rangle +\langle \sigma_j^y \sigma_{j+1}^x\rangle).
$

\section{Results}
\label{sec:Results}

\subsection{Numerical Evaluation of Dynamics}
\label{sec:numerics}

\begin{figure}[tbp]
    \begin{overpic}[width=\linewidth]{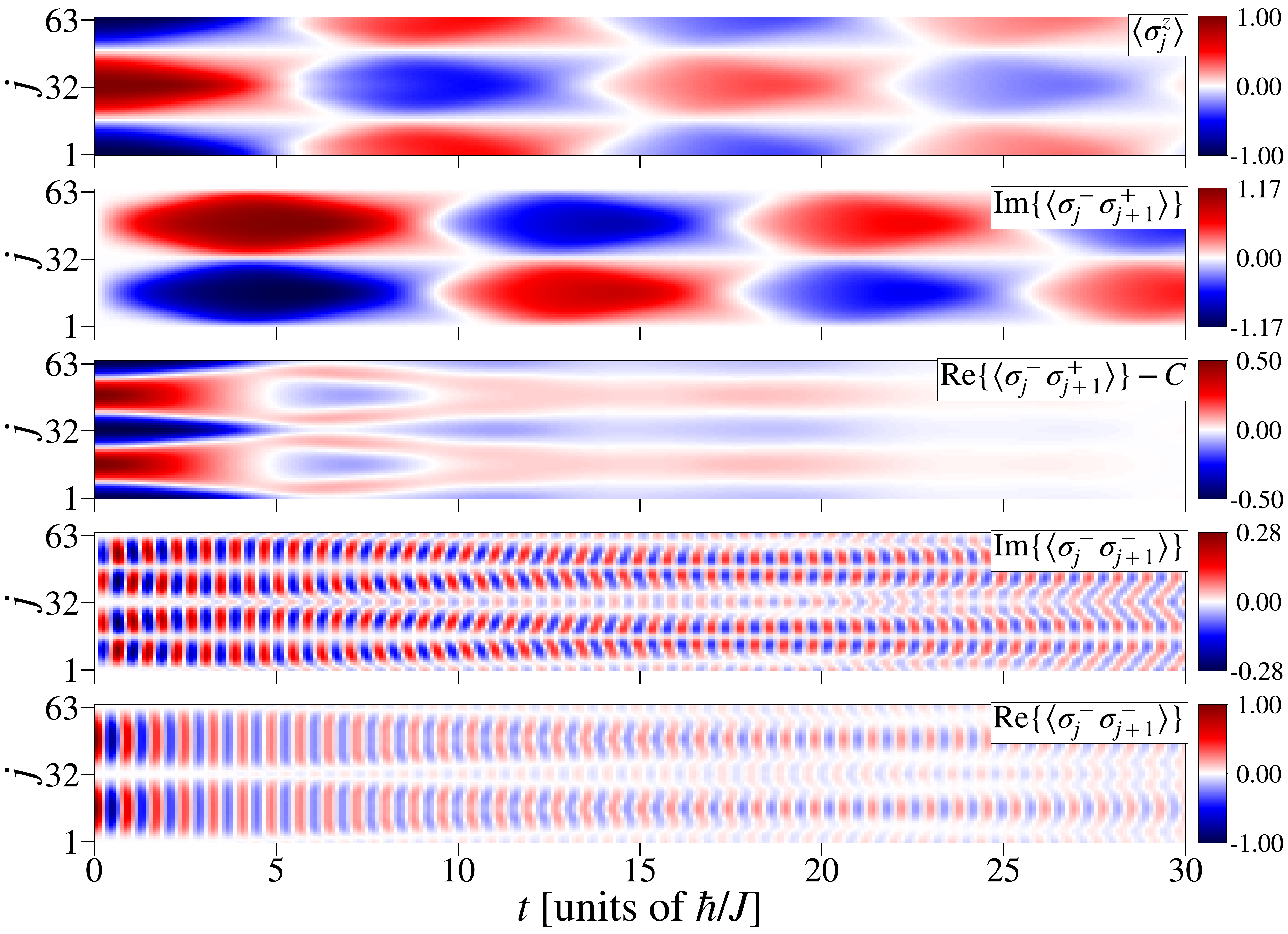}
    \put(-2,70){(a)}
    \put(-2,56){(b)}
    \put(-2,42){(c)}
    \put(-2,30){(d)}
    \put(-2,16){(e)}
    \end{overpic}
    \caption{(Color online) Exact quantum calculation of the time dependence of spin correlations: (a) $\expval{\sigma_j^z}$, (b) $\Im\{ E^{-+}_j\} = \Im \{\expval{\sigma_j^- \sigma_{j+1}^+ }\}$, (c) $\Re \{E^{-+}_j\}  = \Re \{\expval{\sigma_j^- \sigma_{j+1}^+ }\}$, 
    (d)  $\Im{E^{--}_j} = \Im{\expval{\sigma_j^- \sigma_{j+1}^-}}$, (e) $\Re{E^{--}_j} = \Re \{\expval{\sigma_j^- \sigma_{j+1}^-}\} $.  Time $t$ is measured in units of $\hbar/J$ and the position $j$ is dimensionless.  The spin variable $\sigma_j$ is dimensionless, as are the correlation functions.
    The spin helix has wavelength $\lambda=64$, phase $\phi=0$, and repeats periodically in $\lambda$ along the chain. In (c), note that $C = \frac{1}{\lambda} \sum_j \Re \{E_j^{-+}(t)\} $ is conserved for all time and is subtracted off from the plot.}
    \label{fig:longwavelength}
\end{figure}

Figures~\ref{fig:longwavelength} and \ref{fig:Expvals} show a typical time series for the correlation functions introduced in Sec.~\ref{subsec:SpinObservables}.  Figure~\ref{fig:longwavelength} shows the case where the helix wavelength is large, $\lambda=64$, while Fig.~\ref{fig:Expvals} shows $\lambda=8$, which is comparable to the experimental values in Ref.~\cite{JepsenNature2020}. In our simulations, we use a long chain of length $L = 500$ and show results from a slice of length $\lambda = 64$ or $\lambda = 8$ near the middle of the chain. The influence of the boundaries propagates inward at a finite velocity and during the duration of our simulation does not reach the  visible region.

One sees that $\langle \sigma_j^z\rangle$ and $E_j^{-+}$ [Figs.~\ref{fig:longwavelength}(a)--\ref{fig:longwavelength}(c) and \ref{fig:Expvals}(a)--\ref{fig:Expvals}(c)] evolve on a slower timescale than $E_j^{--}$ [Figs.~\ref{fig:longwavelength}(d),\ref{fig:longwavelength}(e) and \ref{fig:Expvals}(d),\ref{fig:Expvals}(e)], and that this separation of timescales grows with $\lambda$.  In Sec.~\ref{timescales}, we give a semiclassical argument which explains this behavior.

At $t=0$, the spins lie in the $x$-$z$ plane  and $\langle \sigma_j^z\rangle$ oscillates in space.  The contrast of this oscillation drops with time, nearly vanishing at  $t\sim 6\hbar/J$  in Fig.~\ref{fig:longwavelength} and at $t\sim 0.8 \hbar /J$ in Fig.~\ref{fig:Expvals}.  At these times, one sees large spatial oscillations in $\Im\{E_j^{-+}\}$, corresponding to patches where the in-plane components of the spins are twisting in different  directions.  Over time this pattern  repeats, with the spatial contrast of $\expval{\sigma_j^z}$ and $\Im{E_j^{-+}}$ oscillating out of phase. This is further examined in Sec.~\ref{sec:Contrasts}.

On a similar timescale to the oscillation of $\expval{\sigma_j^z}$, $\Re\{E_j^{-+}\}$, which is proportional to the local energy density, decays to a spatially uniform pattern.  This can be interpreted  as evidence of energy diffusion. 

The real and imaginary parts of $E_j^{--}$ oscillate  with  a period which is slightly smaller than $\hbar/J$.  Our physical interpretation of these patterns comes from comparing a uniform pattern of quantum versus classical spins,  where $E_j^{--}\sim (\sigma^x-i \sigma^y)^2$ tells us about the local orientation of the in-plane component of the spins. This interpretation is discussed in Sec.~\ref{sec:uniform}.
At short times in Fig.~\ref{fig:longwavelength}(d), the imaginary part of $E^{--}$ (which can be interpreted as $2\sigma^x \sigma^y$) vanishes every quarter wavelength. Between these nodes it alternates between a $+ - - +$ or $- + + -$ sign pattern. In Fig.~\ref{fig:longwavelength}(e), the real part (which can be interpreted as $(\sigma^x)^2-(\sigma^y)^2$) instead has nodes at half wavelengths, and at a given time is either positive or negative. Such a pattern suggests that the in-plane component of the spins are slowly twisting about the $z$ axis, rotating by $\pi$ radians over half a wavelength, and reversing direction for the next period. Identical features are seen in the semiclassical model at times less than $\hbar/4J$ (Figs.~\ref{fig:classical}(d) and \ref{fig:classical}(e)). The primary difference is that in the quantum calculation the pattern reverses periodically, while in the semiclassical one the twisting continues.
At longer times, the pattern is slightly more complicated, with some retrograde motion.  This complicated pattern is seen for both $\lambda=64$ and $\lambda=8$, but it is certainly clearer in  Fig.~\ref{fig:longwavelength}.

In the remainder of this paper, we explain the above features and compare our results to the experiment of Ref.~\cite{JepsenNature2020} in Sec.~\ref{sec:ExpComp}.

\subsection{Uniform Textures \& Long-Wavelength Approximation}\label{sec:uniform}
Key insights into the behavior in Sec.~\ref{sec:numerics} come from considering the uniform case,
where
each spin is initially tilted the same angle $\theta$ from the negative-$\hat z$ direction: $A_{j,\downarrow}=\cos(\theta/2)$, $A_{j,\uparrow}=\sin(\theta/2)$.  This can be thought of as a helix with $\lambda \to \infty$.
Evaluating the sums in Eq.~\eqref{eq:E-+Expression} and \eqref{eq:E--Expression} yields (see Appendix \ref{sec:LongWavelengthAppendix})
\begin{align}\label{unifemp}
E^{-+}(t)&=  \sin^2\theta,\\
E^{--}(t) &=
\sin^2\theta \sum_{l=0}^\infty
(-i)^l \cos^l(\theta) (J_l(8 J t)-J_{l+2}(8 J t)). \label{eq:Uniform--}
\end{align}
The latter expression can also be written in terms of the derivatives of Bessel functions, using the identity $2J^\prime_{l+1}(8Jt)=J_{l}(8Jt) - J_{l+2}(8Jt)$.

The correlation function $E^{-+}$ is time independent:  $\Re\{E^{-+}\}$ is proportional to the local energy density, which is conserved.
The imaginary part of $E^{-+}$ is chiral, measuring an in-plane twisting, and will vanish for all time.  

Equation (\ref{eq:Uniform--}), which describes the behavior of $E^{--}$, is not particularly transparent.    To reveal the underlying structure, we take the long-time limit and find
\begin{align}
    E^{--}(t) 
    &= 
    \frac{1}{\sqrt{\pi J t}}
    \left( \Lambda_\uparrow e^{i 2 \Omega_{\rm q} t } +
    \Lambda_\downarrow e^{-i 2 \Omega_{\rm q} t}
    \right)+{\cal O}(t^{-3/2}),
    \label{eq:LongTimeQMLW}
\end{align}
where $\Lambda_\uparrow=\sin^2(\theta/2) e^{-i\pi/4}$, $\Lambda_\downarrow=\cos^2(\theta/2) e^{i\pi/4}$, and the precession frequency is $\Omega_{\rm q} = 4J$.
The terms $e^{\pm i 2\Omega_q t}$ are suggestive of precession about the $z$ axis: if one takes $\sigma^-= \sigma^x-i\sigma^y \approx e^{i\Omega t}$, then
$E^{--}=\langle \sigma_j^- \sigma_{j+1}^-\rangle \sim e^{2 i \Omega t}$.
Thus it is tempting to interpret
Eq.~(\ref{eq:LongTimeQMLW}) as a quantum superposition of clockwise and counter-clockwise precession.  Note that the magnitude of the precession rate $\Omega_{\rm q}$ is independent of the tilt angle $\theta$.
This discreteness  
aligns  with the idea that each spin is ``measuring" the state of its neighbors and precesses about the resulting field that it sees.  In a given basis, the spins take on only a discrete set of values, leading to only a discrete set of frequencies. Recognizing that $\abs{\Lambda_\uparrow} = A_{j,\uparrow}^2$ and $\abs{\Lambda_\downarrow} = A_{j,\downarrow}^2$ is also suggestive of this picture of projectively measuring the two neighboring spins. Finally, the overall $t^{-1/2}$ decay is related to the build-up of correlations between neighboring sites.

We contrast this behavior to 
a semiclassical approximation, where the spin-wavefunction is constrained to take a product form.  As detailed in
Appendix~\ref{sec:SemiclassicalAppendix}, in this semiclassical approach, the spins uniformly precess about the $z$-axis.  The frequency, $\Omega_{\rm sc}=4 J \cos(\theta)$, is typically smaller than $\Omega_{\rm q}$ and depends on the tilt angle $\theta$.  The local field seen by each spin is not quantized in the semiclassical model. 

We next use Eqs.~\eqref{unifemp} and \eqref{eq:Uniform--} to model the dynamics of spin helices  in the long-wavelength limit.  Treating the spin texture as \textit{locally} homogeneous, we simply replace the $\theta$ in Eqs.~\eqref{unifemp} and \eqref{eq:Uniform--} with $\theta_j=Qj+\phi$.  While such an approximation fails to give  any insight  into $\expval{\sigma_j^z}$ or $E_j^{-+}$, it is revealing in regards to
$E^{--}_j$.  In particular, since $\Omega_{\rm q}$ does not depend on $\theta$, the spin dynamics are approximately periodic.  Both the period and the power-law decay match what is seen in Figs.~\ref{fig:longwavelength} and \ref{fig:Expvals}.

\subsection{Separation of Timescales for In-Plane and Out-of-Plane Dynamics}\label{timescales}

The separation of timescales in Sec.~\ref{sec:numerics} is also present in the semiclassical results shown in Appendix \ref{sec:SemiclassicalAppendix} and hence can be understood semiclassically. In this picture, the spin at site $j$ precesses about an effective field  $\vec{H}_j = 2J(\vec{\sigma}_{j-1} + \vec{\sigma}_{j+1})_{\perp}$, where $\perp$ indicates that only the in-plane component is taken. This is a feature of the $XX$ model, where the spin-spin coupling only involves the in-plane components.

Within the long-wavelength limit of a spin helix, the first-order approximation to the effective field comes from taking $\vec{\sigma}_{j-1} + \vec{\sigma}_{j+1} \approx 2\vec{\sigma}_j$. The next-order correction comes from considering that neighboring spins are \textit{mostly} aligned, but twist in the direction of the helix's winding. Hence, $\vec{\sigma}_{j-1} + \vec{\sigma}_{j+1} \approx 2\vec{\sigma}_j + \frac{d^2}{2} \frac{\partial^2 \vec{\sigma}_j}{\partial x^2}$ for spacing $d$ between spins, and 
\begin{equation}
    \partial_t \vec{\sigma_j} \approx 4J(\vec{\sigma}_j \times \vec{\sigma}_{j,\perp}) + 2J \left(\vec{\sigma}_j \times \frac{d^2}{2} \frac{\partial^2 \vec{\sigma}_j}{\partial x^2}\bigg|_{\perp}\right).\label{sclw}
\end{equation}
The first term determines twisting in the plane and has a timescale of $\Omega \sim J$, which was seen earlier for the precessional frequency. The second term is a correction which determines twisting away from the winding direction. This gives out-of-plane dynamics with a frequency $\omega \sim \frac{J}{\lambda^2}$, which is suppressed by a factor of $\lambda^2$ relative to the in-plane dynamics.

\subsection{Contrasts}
\label{sec:Contrasts}

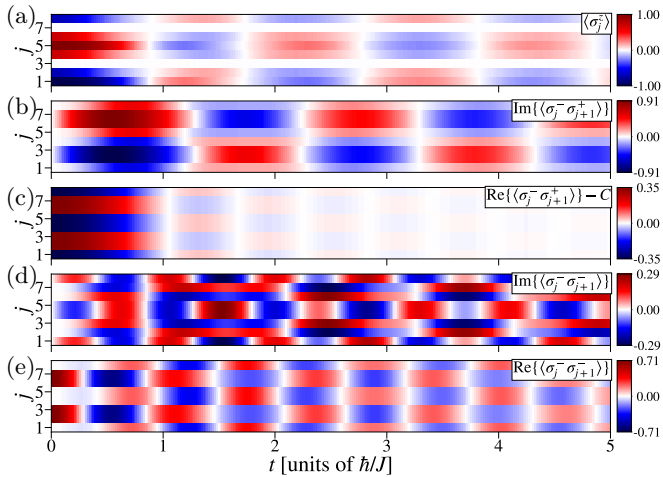
\begin{figure}[tbp]
    \begin{overpic}[width=\linewidth]{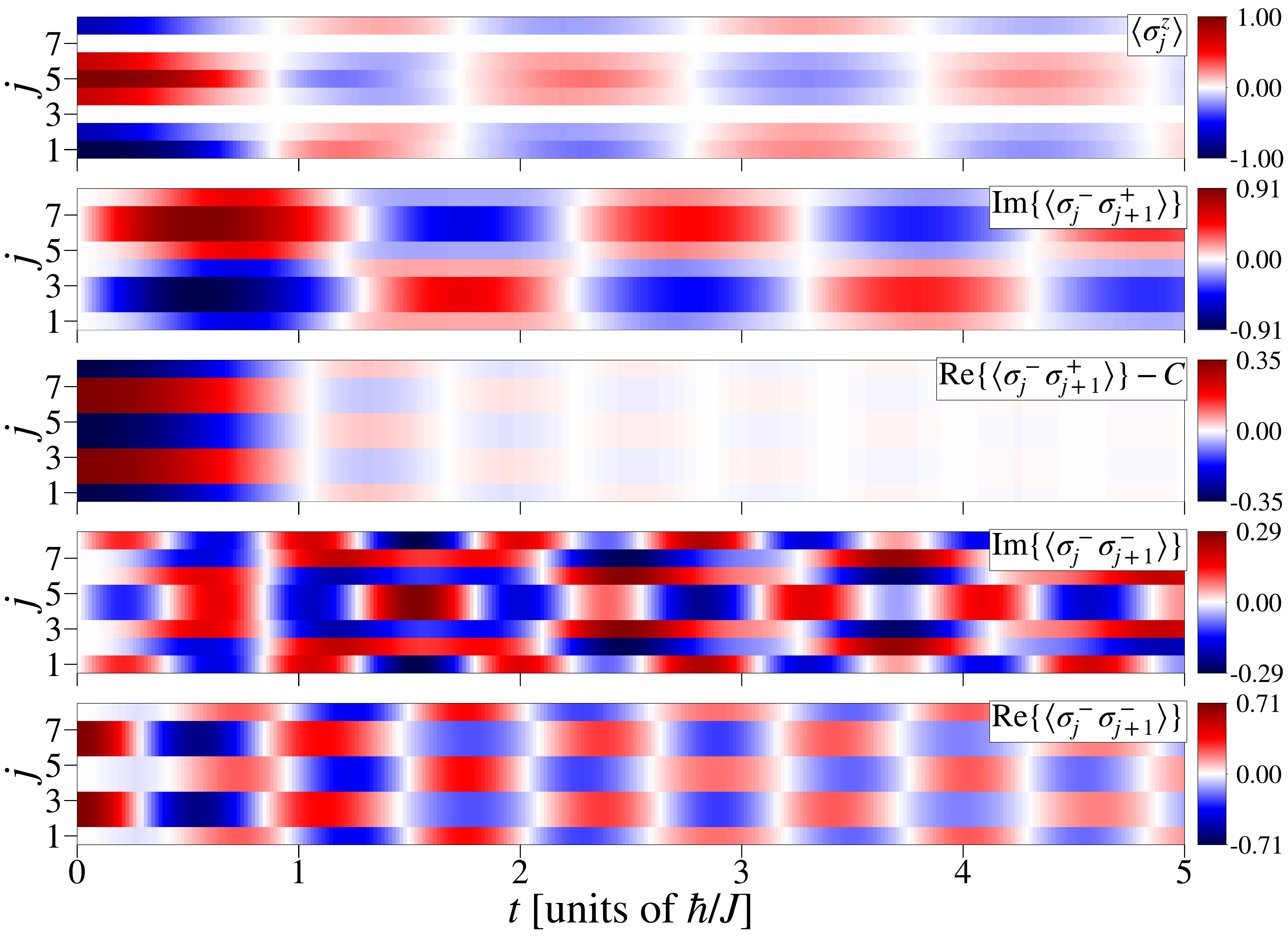}
    \put(-1,70){(a)}
    \put(-1,56){(b)}
    \put(-1,42){(c)}
    \put(-1,30){(d)}
    \put(-1,16){(e)}
    \end{overpic}
    \caption{Exact quantum calculation of the same observables as Fig.~\ref{fig:longwavelength}, but with wavelength $\lambda=8$ and phase $\phi=0$.}
    \label{fig:Expvals}
\end{figure}

For all time, the spin patterns are periodic in space with period $\lambda$.  The amplitude of variation falls with time.  We can quantify this decay by looking at the magnitude of the Fourier component with wavevector $Q=2\pi/\lambda$.  For example,
\begin{align}
    C_Q^z(t) &\equiv \frac{1}{\sqrt{L}} \sum_j e^{iQj} \expval{n_j(t)}.
\end{align}
This quantity can be expressed in terms of the Fourier transform of the initial density matrix,
\begin{equation}
    \rho_{kq} = \frac{1}{L} \sum_{l,m} \rho_{lm} e^{ikl} e^{-iqm}.
\end{equation}
Periodicity of the spin configuration implies
$\rho_{l+\lambda,m+\lambda} = \rho_{lm}$,
and hence  $\rho_{kq}$ vanishes unless $k-q = 2\pi n/\lambda$ for integer $n$.  This leads to the simple result
\begin{align}
    C_Q^z(t) &= \sum_m   \bigg(d_m+ (-1)^m d_m^*\bigg) J_{m}(8J \tau)  \label{eq:ZContrast}
\end{align}
for time-independent coefficients
\begin{equation}
    d_m = \frac{1}{\sqrt{L}} \sum_{j} e^{iQ(m/2-j)} \rho_{j-m,j}
\end{equation}
and $\tau = \sin(Q/2)t$. 
Similarly, the contrast corresponding to $E_j^{-+}(t)
=\langle \sigma_j^-(t)\sigma_{j+1}^+(t) \rangle$ is
\begin{eqnarray}
    C^{-+}_{Q}(t) &=& \frac{1}{\sqrt{L}} \sum_j e^{iQj} \expval{\sigma_j^-(t) \sigma_{j+1}^+(t)}\\\nonumber
   &=& 4\sum_m \bigg(\tilde{d}_{m}
   + (-1)^{m} \tilde{d}_{m}^* \bigg) J_{m}(8J \tau) \label{eq:PMContrast}
\end{eqnarray}
for the coefficients 
\begin{align}
    \tilde{d}_m &=
    e^{-iQ/2} d_{m+1}=
    \frac{1}{\sqrt{L}} \sum_{j} e^{iQ(m/2-j)} \rho_{j-m-1,j}.
\end{align}
For a helix with reflection symmetry across the origin (i.e. $\phi=0$), $C_Q^z(t)$ is real.
Since $\Im{E_j^{-+}(t)}$ evolves on the same timescale as $\langle \sigma_j^z\rangle$, it is convenient to display their contrasts together.
Thus we extract the contrast of $\Im{E_j^{-+}(t)}$ as $C_{Q,i}^{-+}(t) = \frac{1}{2i}\left(C^{-+}_{Q}(t) - (C^{-+}_{-Q}(t))^*\right)$.  This quantity is also real when the density matrix is real, since then $C^{-+}_{-Q}(t) = C^{-+}_Q(t)$.

\begin{figure*}[tbph]
    \centering
    \includegraphics[scale=0.4]{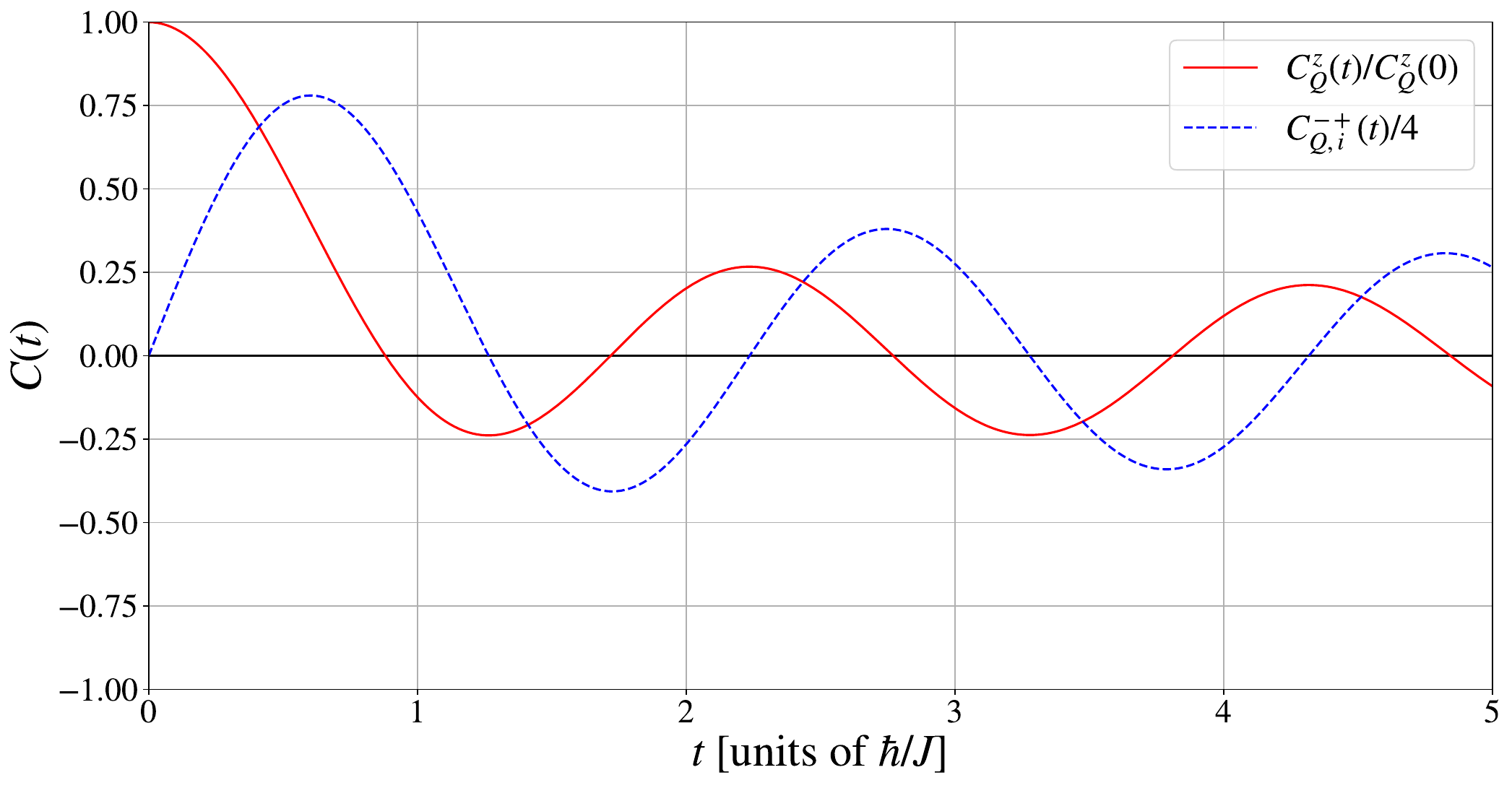}
    \caption{Contrasts $C_Q^z(t)$ (solid red) and $C_{Q,i}^{-+}(t)$ (dashed blue) corresponding to the momentum $Q=2\pi/\lambda$ Fourier components of $\expval{n_j(t)}$ and $\Im{E_j^{-+}(t)} = \Im{\expval{\sigma_j^-(t) \sigma_{j+1}^+(t)}}$, respectively, for a spin helix with wavelength $\lambda = 8$ and phase $\phi =0.$ Both contrasts are rescaled for ease of comparison.}
    \label{fig:Contrasts}
\end{figure*}

Figure~\ref{fig:Contrasts} shows the time dependence of the contrasts $C_Q^z(t)$ and $C_{Q,i}^{-+}(t)$ for the case $\lambda=8$.  
The most prominent feature is an
out-of-phase relationship between these two contrasts. This further corroborates the out-of-phase relationship between $\expval{\sigma_j^z}$ and $\Im\{E_j^{-+}\}$ that was noted in Sec.~\ref{sec:numerics}. This  relationship corresponds to the spins moving in and out of the $x$-$y$ plane.
The second important feature is the relatively slow $t^{-1/2}$ decay, which can be extracted from the asymptotic expressions for the Bessel functions.

\section{Comparison to Experiment}
\label{sec:ExpComp}

The experiment of Jepsen \textit{et al.} \cite{JepsenNature2020} explored spin helices with wavelengths similar to that in Fig.~\ref{fig:Expvals}.  They solely looked at the space and time dependence of $\langle \sigma_j^z\rangle$.  Other spin correlations, however, are accessible: they could apply $\pi/2$ pulses to image the $x$- and $y$-components of the spins.  Such data could be binned and averaged to measure $E_j^{-+}$ or $E_j^{--}$.

The experimentalists calculated the contrast of $\langle \sigma_j^z\rangle$, to be compared with our results depicted in Fig.~\ref{fig:Contrasts}.  They also used numerical techniques to model the spin dynamics.  Their numerical results for the contrast appear to be identical to ours.  

As already highlighted, the contrast oscillates within a decaying envelope.  The experimental oscillation period is very similar to what is expected from theory, both from our calculation and their numerics.  The decay, however, is very different.  Our theory predicts a $t^{-1/2}$ fall-off, while the experiment observes an exponential decay.  Possible sources of this discrepancy include
the presence of empty sites in the optical lattice or inhomogeneities in the magnetic fields which tune interactions in the chain.

The nature of the decay is important: the experimentalists used the $Q$ dependence of the decay time to conclude that the dynamics in the $XX$ model are ballistic.  Given that the exact solution does not yield an exponential decay, this interpretation may be problematic.

More fundamentally, the power-law decay seen in the $XX$ model is a consequence of integrability.  The experimental deviations are indicative of terms which break the underlying symmetries. Moreover, these deviations are not small:  discrepancies appear on a timescale of a few $\hbar/J$, which means the energy scale of the perturbation is comparable to $J$.

In order to clarify the issue, we propose that future experiments study the dynamics of uniformly-tilted spins, as discussed in Sec.~\ref{sec:uniform}.  Many of the dynamical features of spin helices are already present in that simpler setting, but the homogeneity of the initial state will simplify the modeling of various imperfections.

\section{Summary and Outlook}
\label{sec:conclusion}

We studied the dynamics of far-from-equilibrium spin helices in the quantum 1D $XX$ model. By mapping the spins onto noninteracting fermions, we were able to derive exact results for a number of spin correlators.  For long-wavelength helices, we found a separation of timescales between the in-plane and out-of-plane spin dynamics.  
We explained these timescales by
 analyzing a homogeneous spin chain as well as a semiclassical model, which simultaneously exposed the role of quantum effects on the dynamics.  Finally, we compared our results to experiments, finding key differences which highlight an important challenge in quantum simulation:  small imperfections can lead to qualitatively different physics.

The community's newfound ability to experimentally study spin models has great potential to advance our understanding of quantum dynamics.  These possibilities 
include
studies of the dynamics of various spin configurations \cite{PhysRevE.59.4912, Moriya_2019, PhysRevE.69.066103, PhysRevLett.110.060602, RigolReview}  and developments in generalized hydrodynamics \cite{FagottiGHD2016, YoshimuraGHD2016, YoshimuraGHD2017, DubailGHD2019, DubailGHD2020, RigolGHD2021}.  
The possibilities involving far-from-equilibrium physics are particularly rich, with ample opportunities to develop new organizing principles.  These will have an impact on our understanding of natural phenomena, as well as the development of future technology.

\section*{acknowledgement}
This material is based upon work supported by the National Science Foundation under Grant No.  PHY-2110250. We also acknowledge the support of the Natural Sciences and Engineering Research Council of Canada (NSERC) (Ref. No. PGSD-567963-2022).

\appendix
\section{Spin Correlators for Uniform Spin Textures}
\label{sec:LongWavelengthAppendix}

When the spin texture is uniform, the spin correlator $E^{--}_j(t) =4\langle a_{j+1}(t) a_j(t)\rangle$ is independent of the site and can be replaced by its spatial average $E^{--}(t) \equiv 1/L \sum_j E_j^{--}(t)$. This is related to the $p$-wave pairing amplitude,
\begin{equation}
    E^{--}(t)=\frac{4}{L}\sum_k e^{i k} \langle b_k(t) b_{-k}(t) \rangle.
\end{equation}
We substitute in the time dependence $b_k(t)=e^{-i \epsilon(k) t} b_k(0)$ and write $b_k(0)$ in terms of the spatial field operators $a_j=a_j(0)$.  Noting that $\langle a_i a_j\rangle=\langle a_{i-j} a_0\rangle$, we find
\begin{equation}
    E^{--}(t)=\frac{4}{L}\sum_d \langle a_d a_0\rangle \sum_k e^{-i (d-1) k-2i\epsilon(k) t}. \label{eq:--MidStep}
\end{equation}
The sum over $k$ yields Bessel functions.
Finally, we note that $\langle a_0 a_0\rangle=0$ and, for $d\neq 0$, Eq.~\eqref{eq:PairCorMat} becomes
\begin{equation}
    \expval{a_d a_0} = \frac{1}{4}\sin^2(\theta) \cos^{\abs{d}-1}(\theta),
\end{equation}
which falls off exponentially. Separating out the $d>0$ and $d<0$ terms in Eq.~\eqref{eq:--MidStep} yields 
\begin{align}
    E^{--}(t) &= \sin^2(\theta) \sum_{l=0}^{\infty} (-i)^{l} \cos^l(\theta) \left[J_{l}(8Jt)-J_{l+2}(8Jt)\right],
\end{align}
which is Eq.~\eqref{eq:Uniform--}.

Calculating $E^{-+}(t) \equiv 1/L \sum_j E_j^{-+}(t)$ follows a similar logic; the main difference is that the momentum space density $n_k =\langle b_k^\dagger(t) b_k(t)\rangle$ appears instead of the $p$-wave pairing amplitude. Since $n_k$ is independent of time, so is $E^{-+}$. Hence, $E^{-+}(t) = E^{-+}(0)$, or
\begin{align}
    E^{-+}(t) &= 4\abs{A_{\uparrow} A_{\downarrow}}^2 = \sin^2(\theta).
\end{align}

As argued in the main text, we can use these homogeneous results to model a long-wavelength helix.  We take the homogeneous expressions for the correlators, but replace $\theta\to\theta_j=Q j+\phi$, arriving at
\begin{align}
    E_j^{--}(t) &\approx \sin^2(\theta_j) \sum_{l=0}^{\infty} (-i)^{l} \cos^l(\theta_j) \left[J_{l}(8Jt)-J_{l+2}(8Jt)\right], \\
    E_j^{-+}(t) &\approx \sin^2(\theta_j).
\end{align}


\section{Semiclassical Spin Dynamics}
\label{sec:SemiclassicalAppendix}
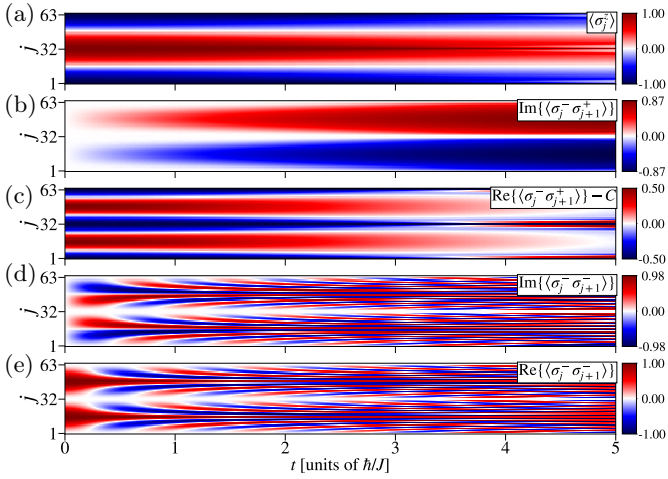
\begin{figure}[tb]
    \begin{overpic}[width=\linewidth]{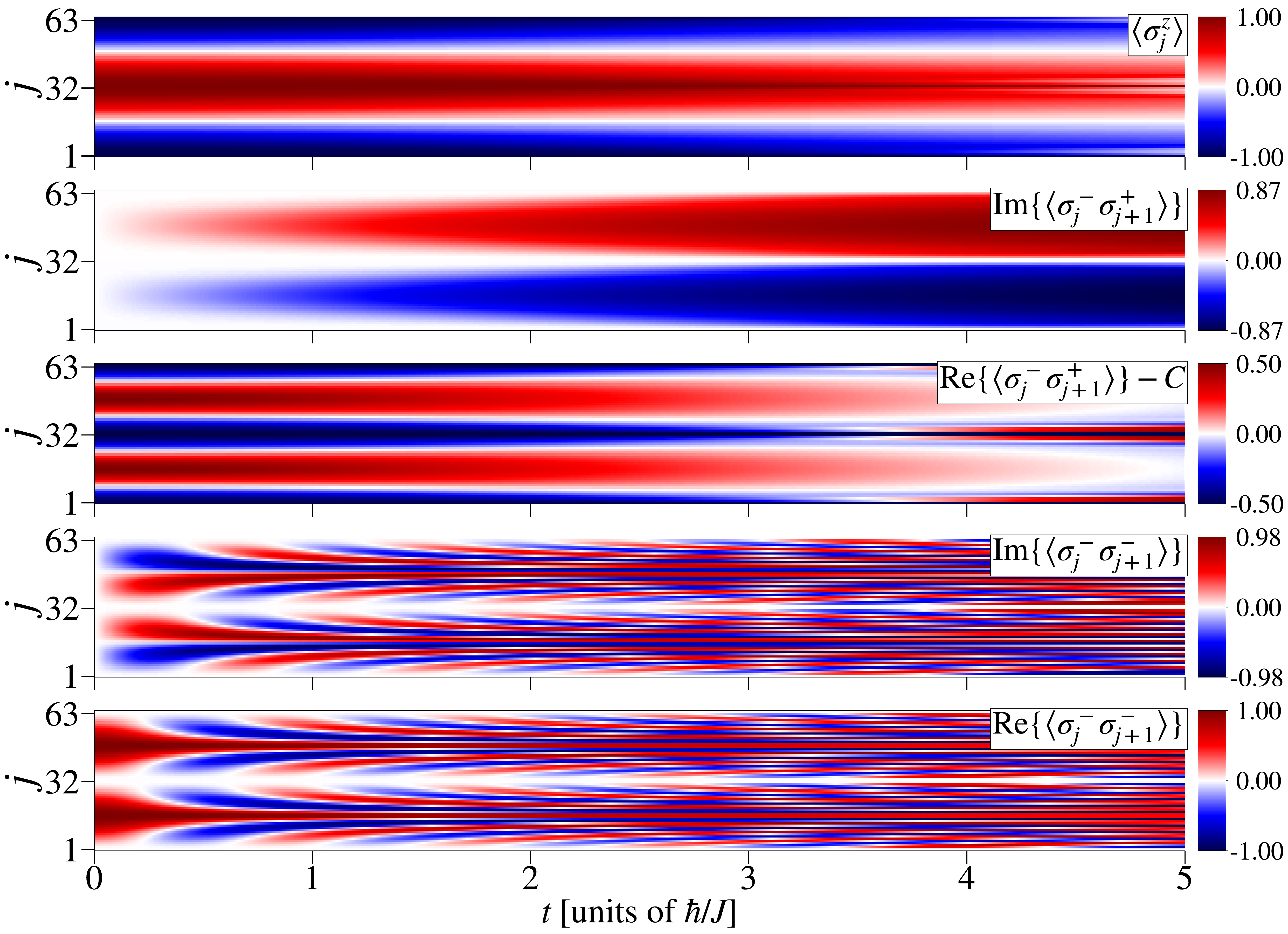}
    \put(-2,70){(a)}
    \put(-2,56){(b)}
    \put(-2,42){(c)}
    \put(-2,30){(d)}
    \put(-2,16){(e)}
    \end{overpic}
    \caption{Semiclassical calculation  of the same observables as Fig.~\ref{fig:longwavelength}, again with wavelength $\lambda=64$ and phase $\phi=0$.}
    \label{fig:classical}
\end{figure}

The Schr\"odinger equation is equivalent to extremizing the action
\begin{equation}
    S = \int dt \left[i \bra{\psi} \partial_t \ket{\psi} - \bra{\psi} \mathcal{H} \ket{\psi} \right]. \label{action}
\end{equation}
We take the product state ansatz,
$\ket{\psi} = \prod_j \ket{\psi}_j$, where $\ket{\psi}_j = u_j(t) \ket{\uparrow}_j + v_j(t)\ket{\downarrow}_j$ with $|u_j(t)|^2+|v_j(t)|^2=1$.  
The semiclassical equations of motion are found by extremizing Eq.~\eqref{action} with respect to $u_j(t)$ and $v_j(t)$.

In particular, we take
\begin{equation}
    \ham = \sum_j \sum_{\mu \nu} J^{\mu \nu} \sigma_j^\mu \sigma_{j+1}^\nu.
\end{equation}
The equations of motion for the spin degrees of freedom are then found to be
\begin{align}
    \partial_t \expval{\sigma_j^\gamma} &= \partial_t u_j \frac{\partial \expval{\sigma_j^\gamma}}{\partial u_j} + \partial_t v_j \frac{\partial \expval{\sigma_j^\gamma}}{\partial v_j} + \textrm{h.c.} \\
    &\equiv  \sum_{\mu \tau} \epsilon_{\gamma \mu \tau} H_j^\mu \expval{\sigma_j^\tau},
\end{align}
where the mean field $H_j^\mu$ is 
\begin{equation}
	H_j^\mu \equiv 2 \sum_\nu J^{\mu \nu} (\expval{\sigma_{j+1}^\nu} + \expval{\sigma_{j-1}^\nu}).
\end{equation}
These equations of motion can also be written as $\partial_t \expval{\vec{\sigma}_j} =  \expval{\vec{\sigma}_j} \times \vec{H}_j$, which is the well-known Landau-Lifshitz equation \cite{SpinDynamicsBook}:
each spin simply precesses about an effective field $\vec{H}_j$, which depends on the direction of its neighbors and the anisotropy of the interaction coupling. 
In the case of the $XX$ model, the mean field $H_j^\mu$ has $H^z_j=0$ and in-plane components 
\begin{eqnarray}
H^x_j=2J (\expval{\sigma_{j-1}^x}+\expval{\sigma_{j+1}^x}),\\
H^y_j=2J (\expval{\sigma_{j-1}^y}+\expval{\sigma_{j+1}^y}).
\end{eqnarray}
Starting from an initial spin helix, we integrate these equations of motion  using a fourth-order Runge-Kutta algorithm \cite{NewmanComputationalBook}.  As before, we eliminate concerns about boundaries by taking a chain of length $L=500$ and confining our attention to a region far from the boundary.

To understand the role of quantum mechanics, it is necessary to compare the exact quantum dynamics in Fig.~\ref{fig:longwavelength} with the semiclassical results shown in Fig.~\ref{fig:classical}. 
The semiclassical theory captures many features of the quantum spin evolution. In particular, the time dynamics for $\expval{\sigma_j^z}$, $\Im{E_j^{-+}}$, and $\Re{E_j^{-+}}$ are nearly indistinguishable from the quantum dynamics shown in Figs.~\ref{fig:longwavelength}(a)--\ref{fig:longwavelength}(c) until $t \approx 5\hbar/J$. Hence, the same interpretation can be used for both the quantum and semiclassical cases for these three correlators (e.g. the decay of $\Re\{E_j^{-+}\}$ representing energy diffusion). Beyond $t \approx 5\hbar/J$, the semiclassical simulations possess a numerical instability.

The semiclassical theory does not capture the behavior of $E_j^{--}$. To understand the discrepancy, it is useful to again consider a uniform chain as in Sec.~\ref{sec:uniform}, except  with semiclassical spins. Each site is identical, so  $\partial_t \vec{\sigma}= \vec\sigma\times \vec H$ with $\vec{H} = 4J(\sigma^x, \sigma^y, 0)$. We can use that $\vec{\sigma}\times\vec{\sigma}=0$ to write $\vec\sigma\times \vec H = \vec\sigma\times (\vec H-4J\vec{\sigma})=-4 J \sigma^z \vec\sigma\times \hat z$. This means that the semiclassical spins simply precess about the $\hat z$ axis with frequency 
$\Omega_{\rm sc}=4 J \cos(\theta)$.  Consequently, $\sigma^{\pm}=\sin(\theta) e^{\pm i \Omega_{\rm sc} t}$ and
\begin{eqnarray}\label{unifsemipm}
    E^{-+}_{\rm sc}(t)&=& \sin^2(\theta),\\\label{unifsemimm}
    E^{--}_{\rm sc}(t)&=& \sin^2(\theta) e^{-2 i\Omega_{\rm sc} t}.
\end{eqnarray}
Importantly, the precession frequency is a continuous function of the angle $\theta$.
This behavior should be contrasted with the quantum behavior in Eq.~(\ref{eq:LongTimeQMLW}), where regardless of $\theta$ the spins precess with frequency $\pm 4J$.  The natural interpretation is that the effective field is quantized in the full quantum theory.  The semiclassical theory also fails to account for the power-law decay seen in Eq.~(\ref{eq:LongTimeQMLW}).

The semiclassical result for $E_j^{--}$ can be further understood through a long-wavelength approximation. As previously argued, we can approximate the behavior of a 
long-wavelength helix by taking $\theta \to \theta_j = Qj + \phi$. In this approximation, the semiclassical correlation functions become 
\begin{eqnarray}\label{lwlsemipm}
    E^{-+}_{j, \rm sc}(t)&=& \sin^2(\theta_j),\\\label{lwlsemimm}
    E^{--}_{j, \rm sc}(t)&=& \sin^2(\theta_j) e^{-2 i\Omega_{j, \rm sc} t},
\end{eqnarray}
for $\Omega_{j,\rm sc} = 4J \cos(\theta_j)$.  The phase factor $E^{--}_{j, \rm sc}(t)\propto e^{-2 i\Omega_{j, \rm sc} t}$ corresponds to a variable helical pattern in the $x$-$y$ plane:  the pitch increases linearly in time, and the spins can rotate many times about the equator.  By contrast, in  the quantum dynamics, the $x$-$y$ components of the spins never perform a full rotation.

The \textit{full} semiclassical behavior of $E_j^{--}$ in Figs.~\ref{fig:classical}(d) and \ref{fig:classical}(e) is similar to this long-wavelength approximation, given by Eqs.~\eqref{lwlsemipm} and \eqref{lwlsemimm}. The aforementioned planar twisting is clear in Figs.~\ref{fig:classical}(d) and \ref{fig:classical}(e), with the pitch becoming greatest near $t=2 \hbar/J$.  Due to aliasing, the pitch then decreases.  There are some quantitative differences between the long-wavelength and exact semiclassical dynamics, but the most significant difference is the numerical instability which appears at longer times.  Given that the instability is not present in the quantum calculation, we do not explore it further.


%

\end{document}